\documentclass[10pt,journal,compsoc]{IEEEtran}
\ifCLASSOPTIONcompsoc
  \usepackage[nocompress]{cite}
\else
  \usepackage{cite}
\fi
\ifCLASSINFOpdf
\else
\fi

\usepackage{graphicx}
\usepackage{subfig}
\usepackage{multirow}
\usepackage{multicol}
\usepackage{algorithm, algpseudocode}
\usepackage{algorithmicx}

\begin{document}
%
\title{A Dataset Generation Framework for profiling Disassembly attacks using Side-Channel Leakages and Deep Neural Networks}
%
%
%
%

\author{Pouya Narimani,
        Seyed Amin Habibi,
        Mohammad Ali Akhaee
\IEEEcompsocitemizethanks{\IEEEcompsocthanksitem P. Narimani, S.A. Habibi and M.A. Akhaee are with the Department
of Electrical and Computer Engineering, University of Tehran, Tehran, Iran (pouya.narimani@ut.ac.ir, Seyedaminhabibi@ut.ac.ir, akhaee@ut.ac.ir).
\protect\\


}
}

%
%

\markboth{Journal of IEEE Transactions on Computers}%
{}
\makeatletter
\long\def\@IEEEtitleabstractindextextbox#1{\parbox{0.922\textwidth}{#1}}
\makeatother

\IEEEtitleabstractindextext{%
\begin{abstract}
Various studies among side-channel attacks have tried to extract information through leakages from electronic devices to reach the instruction flow of some appliances. However, previous methods highly depend on the resolution of traced data. Obtaining low-noise traces is not always feasible in real attack scenarios. This study proposes two deep models to extract low and high-level features from side-channel traces and classify them to related instructions.
We aim to evaluate the accuracy of a side-channel attack on low-resolution data with a more robust feature extractor thanks to neural networks.
As investigated, instruction flow in real programs is predictable and follows specific distributions. This leads to proposing a LSTM model to estimate these distributions, which could expedite the reverse engineering process and also raise the accuracy.
The proposed model for leakage classification reaches 54.58\% accuracy on average and outperforms other existing methods implemented on our datasets. Also, LSTM model achieves 94.39\% accuracy for instruction prediction on standard implementation of cryptographic algorithms.
\end{abstract}

}

\maketitle

\IEEEdisplaynontitleabstractindextext

%
\IEEEpeerreviewmaketitle

\IEEEraisesectionheading{\section{Introduction}\label{sec:introduction}}

%
%
%
%
\IEEEPARstart{R}{everse} engineering using side-channel analysis has become more critical in hardware security in the last two decades due to its ability to extract information from target hardware. Technically speaking, reverse engineering aims to assign a related label to each side channel trace. This label could be a bit in bit-level attack scenarios, where the target is a single bit, or an instruction in instruction-level attacks, in which the target is an instruction. Extracting a specific bit or instruction relies on the amount of information leaks from the target device. Obtaining high resolution data during profiling phase is required in a bit-level attack because the target is one bit. On the other hand, for instruction-level attacks, the target is to find a sequence of bits representing an instruction for which we are looking. It seems that deploying an instruction-level attack is more practical in low sample rates with lower-cost tools.

Although many statistical methods have been developed in the last two decades to disassemble a target microprocessor using side-channel information, obtaining high accuracy in such attacks with low-resolution data is still an open problem. These methods suffer from a common problem, i.e., generalization
\cite{wu2020attack}.
Deploying a reverse engineering attack using side-channel information includes two phases: the profiling phase and the attack phase. In the first phase, the attacker obtains enough knowledge about the processor of the target device. This phase includes gathering side-channel traces while the device is running some instructions. The amount and quality of leakage in this phase is various depending on the different executing scenarios. In this case, the attacker may use different sequences of instructions to gather a dataset that spans as much as possible situations that might happen. Subsequently, this data is applied to model the leakage of the target device and later be used in real attack scenarios. The generalization of this method is based on multiple factors such as noise level, the pipeline of the target device, and the ability of the model to estimate the behavior of the target device.

Recently, most of the proposed methods aim to solve this barrier by
estimating the leakage and recognizing instructions from their related traces
\cite{quisquater2002automatic, suykens1999least, hospodar2011machine, chari2002template}.
These references evaluate their proposed methodologies in real attack scenarios. However, they have some strict assumptions on their target devices and attack scenarios which lead to severe defects including:
\begin{itemize}
	\item
	Employing high-cost measurement tools (high resolution data)
	\item
	Using minority instructions from the instruction set of the target device
	\item
	Target devices are mostly PIC which have a simple pipeline
\end{itemize}

A few state-of-the-art articles suffer from these critical issues, where most of them not satisfy real attack scenarios or require expensive experimental tools. In this study, we will show that employing a well-designed model for feature extraction could distinguish instructions from their leakage, even in low sample rates.

This study uses convolutional neural networks (CNNs) for feature extraction from side-channel traces and a fully connected (FC) layer for classifying extracted features. Since 2012, CNNs have proved to outperform most of the state-of-the-art algorithms in signal processing
\cite{krizhevsky2012imagenet}.
These networks are robust against shifting and scaling due to the localization they apply to the input, leading to spatial feature extraction. This robustness lets these architectures get more information from traces, which make them appropriate for side channel attacks. 
Training such a CNN requires labeled dataset including traces and related instructions. Therefore, due to the lack of an open-source dataset, a custom one has been gathered in this work. The recorded traces should be gathered from random-order instructions to span all possible sequences of instructions while it should be balanced. We will discuss more about the proposed data gathering mechanisms in section \ref{dataset}.

As mentioned, the proposed dataset should be as general as possible to contain the most plausible types of instruction sequences. However, if the attacker obtains knowledge about the target device code and figures out what type of code is running, then the instruction set could be limited. This limitation lets the model distinguish traces better because of an elimination of some possible labels. Furthermore, instruction flow in some specific programs is predictable which means that instructions do not follow each other randomly as assumed before. In fact, they follow some probability distributions. An algorithm based on Long-Short Term Memory (LSTM) would be proposed to estimate these distributions.
LSTM is a deep neural network which can learn and memorize temporal features in its hidden states. This network is very effective, especially dealing with time series. However, its hidden states increase the number of learning parameters, which imposes more data and more computational power. An instruction frequency analysis model is proposed in the section
\ref{model}
based on LSTM to analyze and estimate the instruction flow in a target device on which a specific code is running.

\subsection{Our Contribution}
We first design a scheme to generate assembly programs, including random instructions and random register values. Then, a customized version of an open-source simulator
is employed to simulate generated random programs and extract run time labels. These labels are then used in a supervised learning procedure to train a deep CNN model on gathered traces. Furthermore, a tool including random instruction generator along with a customized simulator is designed to simplify the profiling phase in performing the side-channel attack (SCA).
Two deep neural network models based on CNN architecture are proposed to extract features from side-channel traces for leakage estimation. Then, a FC layer will be used to assign a label to each trace. After training this network, LSTM model is employed to estimate the probability distribution of instructions for standard implementation of cryptographic algorithms.

The achieved results show that using CNN for feature extraction and classification of side-channel traces have more generalization and outperform state-of-the-art reverse engineering models especially for low-resolution traces.
Besides, LSTM model could reduce the possible instructions in the training phase when the attacker possesses information about the type of program running on the target device.
Finally, a discussion about the proposed method and its practicability is presented. Also a table including the instruction recognition probability is proposed, which determines the leakage of each instruction.

\section{Related Works}
\begin{figure*}[t]
	\begin{center}
		\includegraphics[width=11cm]{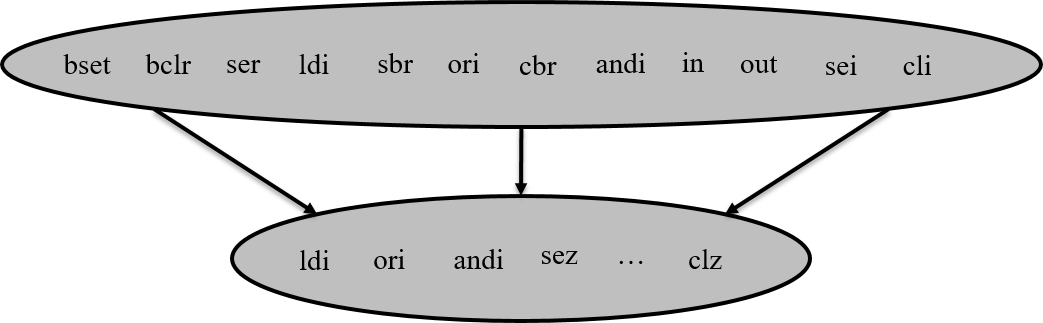}
	\end{center}
	\caption{Instruction mapping between spaces by compiler: The mapping that is translated by compiler is not one-to-one}
	\label{fig:mapping}
\end{figure*}

Several researchers around the world have tried to use side-channel analysis to extract information from devices
\cite{chari2002template, kocher1999differential, brier2004correlation, schindler2005stochastic}.
This information could be employed to extract a hidden key for a cryptographic algorithm or to reverse the whole procedure of the device.
A SCA could be modeled as a classification problem in which the classification is done to extract a hidden key or for instructions detection. Since SCA can be interpreted as assigning a label or instruction to a specific trace, the machine learning classifier can be used as a suitable solution.

From key extraction point of view,
Quisquater et al. employed self organizing map (SOM) for instruction recognition of a de-packaged processor
\cite{quisquater2002automatic}.
Later, the least squares support vector machine (LS-SVM)
\cite{suykens1999least}
is exploited to extract the key of an advanced encryption standard (AES) with no countermeasures
\cite{hospodar2011machine}.
Their achieved results show that LS-SVM with RBF kernel could outperform template attack (TA)
\cite{chari2002template}
on their dataset.
Lerman et al. evaluated different machine learning methods, including SOM, SVM, and random forest(RF) on the implementations of triple data encryption standard (3-DES) and RSA-512 on FPGA
\cite{lerman2014power}.
They later provided results of high dimension data in the profiling phase and compared TA with RF and SVM 
\cite{lerman2015template}.
They indicated that TA outperforms if points of interest could be recognized for data. However, RF can result in better for high dimensional traces.
In \cite{picek2017side}, 
it is also indicated the superiority of machine learning techniques compared to TA while dealing with limited traces.
Thereafter, Cristiani et al. introduced a mutual information estimator based on multi-layer perceptron (MLP) to extract features between input bits and traces
\cite{cristiani2020leakage}.
The proposed method was evaluated with AES implementation on a PIC microcontroller while electromagnetic (EM) traces was used from multiple probe positions.
In 2017, ELMO (leakage simulator for the ARM Cortex-M0) was developed to model second-order bit-interaction terms
\cite{mccann2017towards}.
This work was proposed based on the linear regression-based leakage modeling for ARM Cortex-M0 and Cortex-M4
\cite{mccann2016elmo}.
However, it is claimed that ELMO could not second-order leakages as the original paper stated
\cite{gao2020share}.
As the reviewed papers discussed, machine learning techniques outperform classic methods such as TA, and they are more capable of dealing with high dimensional data
\cite{heuser2012new}.

Extracting the hidden key of a cryptographic function is not the sole approach in side-channel analysis. The leaked information could be employed for reverse engineering and disassembling the bit flow of the whole program. ML techniques are widely used in reverse engineering problems. 
In bit-level approaches, Cristiani et al. presented a disassembler utilizing EM radiations of the target device to extract the entire processed bits
\cite{cristiani2019bit}.	
Their success rate on a PIC microcontroller was 99.41\%. Nonetheless, deploying such attack requires precise measurement devices to fix the probe on desired positions and then record the traces.
Among instruction-level attack scenarios, a multivariate normal model was employed to estimate the templates of instruction classes
\cite{eisenbarth2010building}.
The authors have utilized principal component analysis (PCA)
\cite{jolliffe2005principal}
and the linear discriminant analysis (LDA) for dimensionality reduction and feature selection. The result of this method on 35 instructions from PIC was 70.1\%.
In other work, a SCA was deployed on AVR using the K-nearest neighbor (KNN) for classification and PCA as the feature extractor in
\cite{msgna2014precise}.
Their target subset contained 35 instructions and they reached to near 100\% accuracy. However, target instructions were quite simple and limited, thus recognizing instructions in real codes which contain more than 35 instructions is not possible with this method.
Strobel et al. designed an attack based on the KNN algorithm for classification and polychotomous LDA
\cite{friedman1996another}
for dimensionality reduction
\cite{strobel2015scandalee}. The reported success rate was 96.24\%.
Park et al. employed continues wavelet transform (CWT) to map the traces to the time-frequency domain. Then, they used Kullback–Leibler (KL) divergence for feature extraction, and LDA for dimensionality reduction on the traces
\cite{park2018power}.
Finally, they obtained 99.03\% success rate using quadratic discriminant analysis (QDA). Their target device was AVR, and they have used 112 instructions for the profiling and the attacking phase. However, this method could not reach to a good accuracy on our dataset because of the high sampling resolution of their dateset. It should be noted that 315 samples exist in each instruction on their dataset while in the proposed dataset it is 125 samples. Note that some essential assumptions are required for the performance of ML techniques to deploy SCA, such as having high resolution of sampling on side-channel traces and low noise setup for trace recording. In this article, these constraints have been relaxed, and the model is proposed to perform well under desired circumstances thanks to generalization of CNN model.

The rest of this paper is organized as follows. Section
\ref{dataset}
includes the dataset gathering procedure, customizing simavr, and designing AVR random instruction generator and simulator for side-channel attacks (ARCSim). In section
\ref{model}
, the proposed CNN architectures based on two different models are explained. Then, LSTM model and the training mechanism to estimate the probabilities of instructions occurrences are expressed. Section
\ref{Exper}
contains experimental results for two proposed models, and finally, the achieved results and methodologies are discussed in section
\ref{Con}.

\section{Dataset}\label{dataset}
Since dataset has a critical role in data driven and supervised models in the profiling phase, we explain the proposed approach to gather a labeled dataset.

\subsection{Random Instruction Generation}
For a typical SCA, the leakage estimation of the target device is the most significant phase which forms the baseline of the attack phase. Hence, it is crucial to stimulate all possible states of the target device in the profiling phase in order to obtain a rich dataset. Sample programs that are using in the profiling phase should be diverse, and run-time conditions have to be considered. Some of the main challenges are as below:
\begin{itemize}
	\item
	Jump or branch to undefined memory addresses
	\item
	Covering all possible instructions in AVR instruction set
	\item
	Implementing subroutines with unknown number of instructions and subroutines inside subroutines
\end{itemize}
To satisfy these conditions, a random instruction generator (RIG) for AVR microcontrollers is proposed which is inspired by the program counter (PC) and the stack concept in AVR architecture. The proposed instruction generator can produce entirely random assembly instructions which are implementable on AVR. Besides, all register values that may be read or written in run time are filled with random numbers. The procedure to design this tool is shown in Algorithm
\ref{alg:RIG}.
\begin{algorithm}
	\caption{Random Instruction Generation}\label{alg:RIG}
	\begin{algorithmic}[1]
		\Require $n$ \Comment{Number of instructions}
		\Require $stack.put(0)$ \Comment{Put the first line address on stack}
		\Require $pc \gets 0$ \Comment{program counter}
		\While{$pc < n$}
		\State $inst \gets new\ instruction$
		\State $pc \gets pc + 1$
		\If{$argument\ required$} \Comment{registers, register values, branch address}
		\If{$argument\ uses X,Y,\ or\ Z\ pointer$}
		\State $X,Y,\ or\ Z\ pointer \gets random\ value$
		\EndIf
		\State $argument \gets random\ value$
		\EndIf
		\If{$(inst=jump)\ or\ (inst=branch)$}
		\State $limit\ arguments\ inside\ memory$
		\EndIf
		\If{$inst=subroutine$}
		\State $stack.put(current\ address)$ \Comment{Put a value on stack}
		\EndIf
		\If{$inst=return$}
		\State $pc \gets stack[-1]$ \Comment{Last value of stack}
		\State $stack.pop()$ \Comment{Delete the last value of stack}
		\EndIf
		\EndWhile
	\end{algorithmic}
\end{algorithm}

With respect to the mentioned conditions, the RIG has been developed in Python including all instructions from AVR instruction set. This tool could support all AVR cores, but in our study we have implemented five cores:
\begin{itemize}
	\item
	ATmega8
	\item
	ATmega16
	\item
	ATmega32
	\item
	ATmega128
	\item
	ATmega328
\end{itemize}
The RIG is designed modular, thus extending it to support other AVR cores is straightforward. To extend the RIG to a new core, the instruction set of desired core and the supported memory addresses are required. Then, by importing and selecting the new instruction set for the related core in the Python source code, it will use this instruction set as the reference set. As all instructions from AVR instruction set are implemented, this tool will use those implementations to generate a new assembly code with respect to selected core memory addresses.

\begin{figure*}[!t]
	\begin{center}
		\includegraphics[width=14cm]{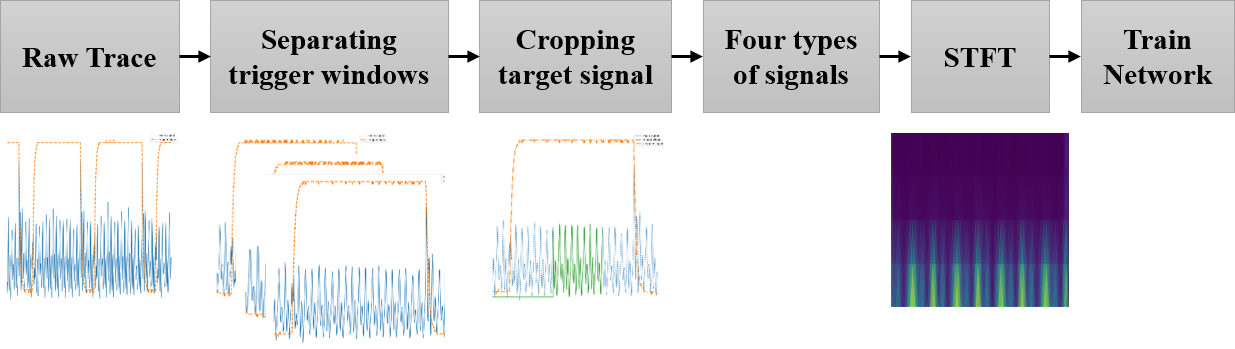}
	\end{center}
	\caption{Preprocessing Procedure: The raw traces first get separated with respect to the trigger signal. Then, we crop the target signal and take the STFT of the cropped signal. Finally, we train our networks by the processed data.}
	\label{fig:PreProcessing}
\end{figure*}

\subsection{Compiling and Programming}
After generating an assembly code including random instructions, AVR-GCC is used to compile the generated assembly file and produce a programmable hex file.
One important issue about AVR-GCC is the mapping between the instructions that are existed in the assembly file and those that are translated to the hex file. The mapping between these two spaces which is translated by the compiler is not one-to-one. In other words, instructions in the assembly file do not necessarily exist in the hex file. Some of these mappings are shown in figure
\ref{fig:mapping}.

Thus, the label space is limited to those existed in the hex file so that some instructions will not exist in the run-time hex file.
After compiling, AVRDUDE
is employed to program the target IC and write the fuse-bits.
AVRDUDE is an AVR downloader and uploader which is exploited on Linux operating system to program the target core.

\subsection{Extracting run-time labels}
As mentioned before, due to using supervised learning algorithms, labels are required to train the proposed model. However, extracting run-time labels for instructions is an undetermined procedure because some instructions are multi-cycle and some of them are one cycle regarding run-time conditions. As a case in point, "brne" could be a one-cycle instruction if a branch is not occurred or a two-cycle instruction if it is branched. Consequently, a simulation of generated hex code is required to obtain run-time labels and simavr is employed
to simulate randomly generated programs.
However, simavr does not export a label file for the simulated program. Therefore, the source code of simavr has been customized to export desired labels in a text file. It is worth mentioning that multi-cycle instructions in the customized version are labeled with indices to separate different cycles of an instruction (e.x. mul\_1 and mul\_2 for two cycles of mul instruction). Thus, all randomly generated programs could be simulated and labeled by a customized version of simavr to estimate the leakage model in the profiling phase.

\subsection{ARCSim}
Combining the RIG with the customized version of the simavr gives the ARCSim in one software.
The ARCSim is designed to simplify the recording of side-channel traces with enough randomness. Furthermore, this software provides instruction labels for the generated programs by employing the customized version of simavr as the simulator. Therefore, using the ARCSim, gathering traces in the profiling phase with related labels is feasible and handy.

\subsection{Recording data}
\begin{figure*}[ht]
	\centering 
	\subfloat[onecycle instruction (clc)]{
		\includegraphics[width=9cm]{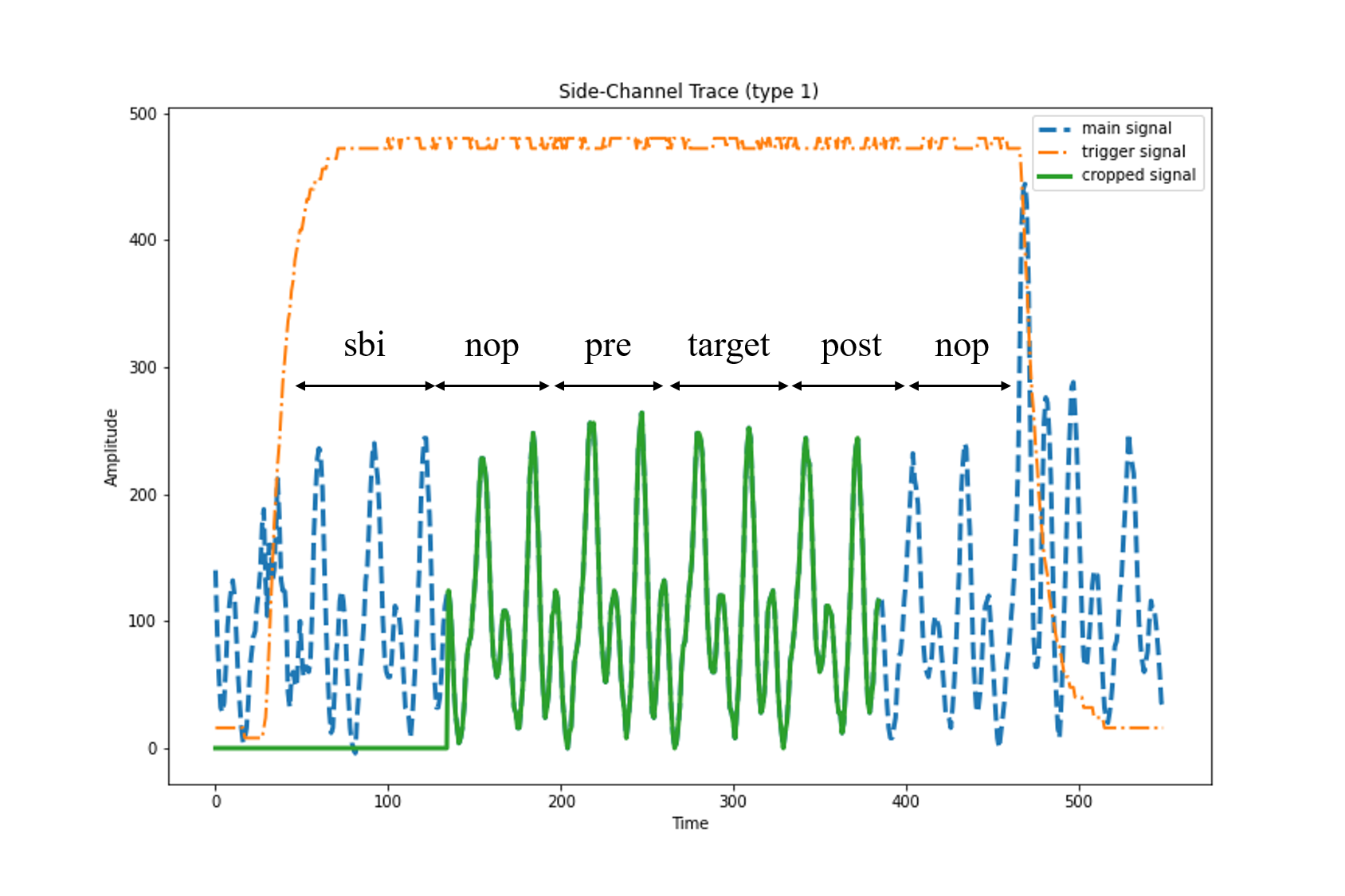}}
	\subfloat[twocycle instruction (muls)]{
		\includegraphics[width=9cm]{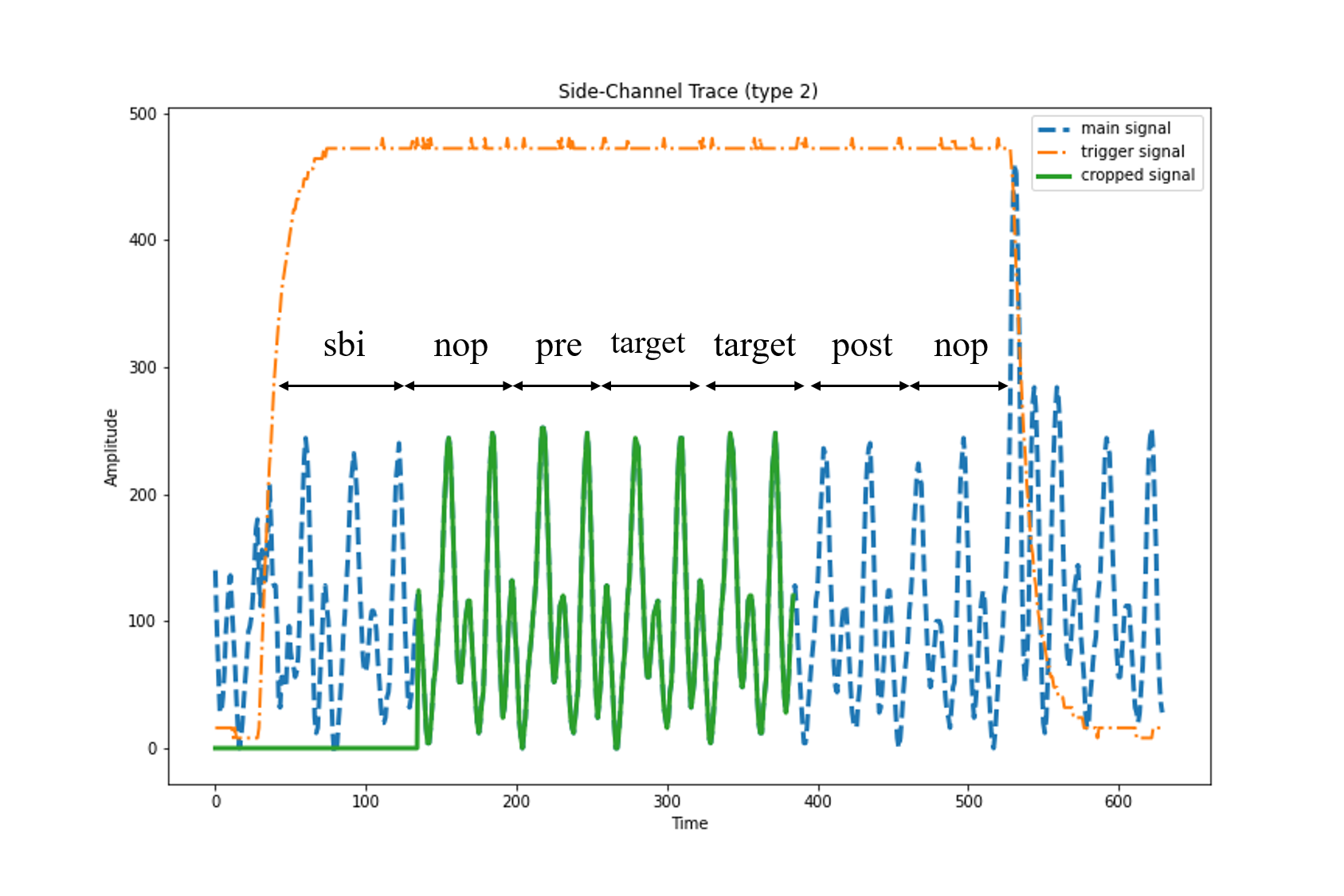}}

	\subfloat[threecycle instruction (ld)]{
		\includegraphics[width=9cm]{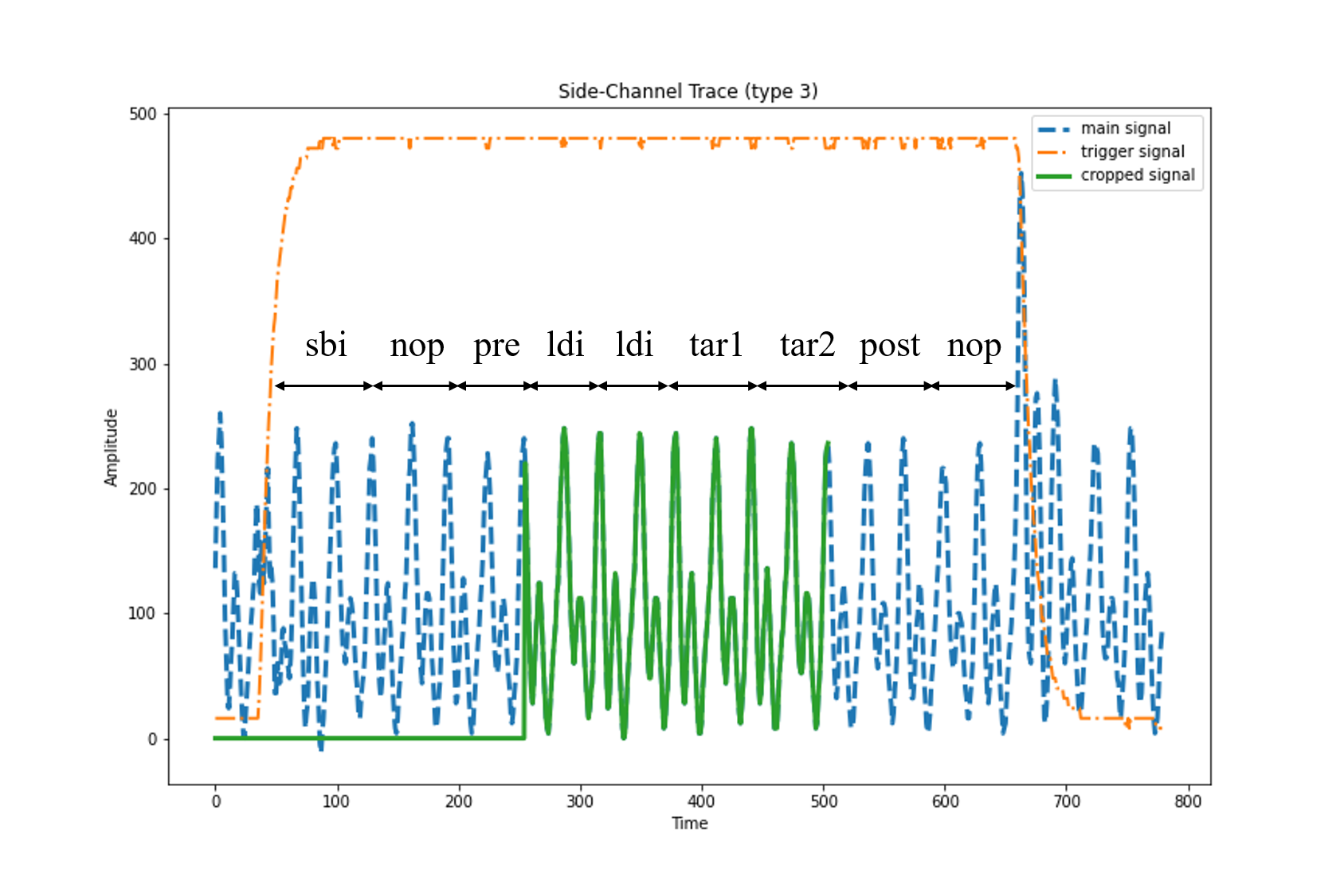}}
	\subfloat[fourcycle instruction (lpm)]{
		\includegraphics[width=9cm]{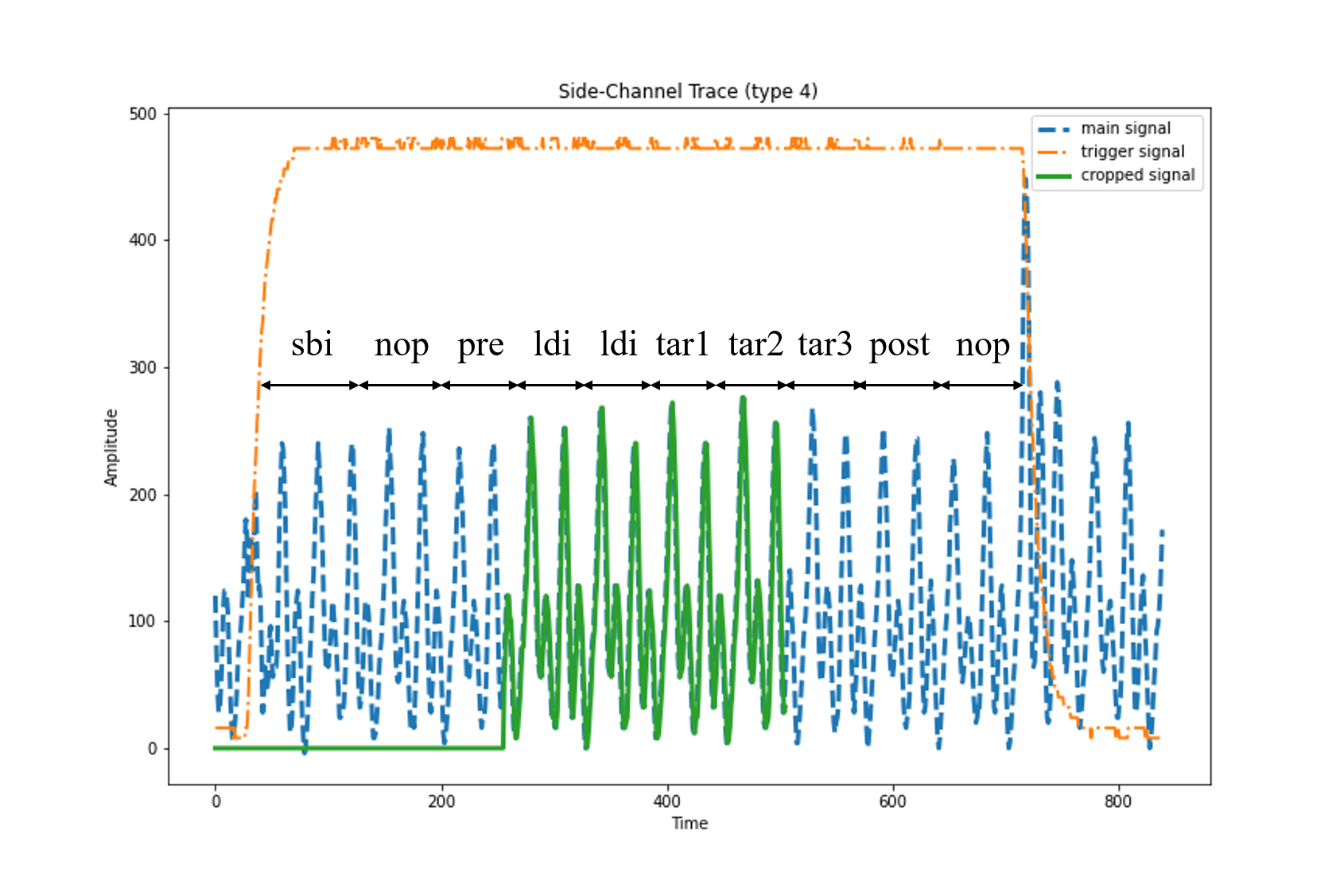}}
	\caption{Cropping instructions: The employed instructions are as follows. "Sbi" is the set bit in I/O register instruction, "nop" is the no operation instruction, "pre" is the random pre-target instruction, "target" is the random target instruction, "post" is the random post target instruction, and "ldi" is the load immediate instruction. Multi-cycle target instructions are splitted to "tar1", "tar2", and "tar3".}
	\label{fig:cropping}
\end{figure*}

After designing and implementing the ARCSim, this tool is used to record side-channel traces on a target device. The framed data structure proposed in
\cite{park2018power}
is employed to record side-channel traces.
Two different datasets have been gathered to train the model and evaluate the proposed methodology. Instructions that manipulate stack value or interact with I/O  ports are discarded in the generation phase because these instructions may cause some run-time pauses, leading to simulation failure.
Power side-channel traces of the target device are recorded by employing the OWON VDS 3102 USB oscilloscope. The sample rate is 1GS/s which is reduced to 500 MS/s when one channel is used as the trace recorder and the other one is applied as the trigger port. Target ATmega runs with an 8 MHz clock frequency, which leads to 125 samples per instruction. 
The descriptions for the gathered datasets are as below:
\begin{itemize}
	\item
	\textbf{Dataset \#1:}
	This dataset includes only one-cycle instructions that ATmega8 can support. It contains 54 instructions from the AVR instruction set, leading to 45 classes due to the mapping problem. Each cropped sample in this dataset consists of 400 sample points with pre-target and post-target instructions.
	\item
	\textbf{Dataset \#2:}
	This dataset is more general and covers 109 instructions from the AVR instruction set, leading to 96 classes due to the mapping problem. Each cropped sample in this dataset consists of 200 sample points with pre-target and post-target instructions.
\end{itemize}

\subsection{Preprocessing}
After gathering traces, some signal processing techniques are required to enhance signal representations before training the proposed model.
They include cropping, alignment, normalization, and short-time Fourier transform (STFT) for better signal representation (Figure \ref{fig:PreProcessing}).
The traces are cropped regarding the cycles of each instruction as shown in Figure \ref{fig:cropping}.
A trigger signal is employed to detect the starting of each frame and consequently aligns the signals. Then, a normalization phase is performed on the traces to scale the values in the interval $[0,1]$. Finally, the STFT is applied to all traces for extracting time-frequency coefficients.

\section{Proposed Model}\label{model}
In this section, two models based on CNN are proposed to estimate the leakage of the target device and learn the distributions of instructions. The first model which has less layers than the second one is proposed to evaluate the capability of CNN in SCA. The second model is proposed to assess the impact of high level features in a typical SCA.
Then, LSTM model is applied to estimate the probability distributions of instructions by knowing previous instructions.

\subsection{Trace classification}\label{classification}
\begin{figure*}[!t]
	\centering 
	\subfloat[VGG-based: This model is named M1 and includes six layers of the VGG architecture.]{
		\includegraphics[width=1\textwidth]{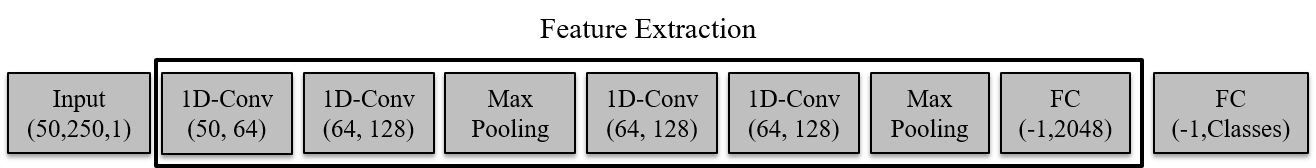}}
	
	\subfloat[ResNet-based: This model is named M2 and it is the customized version of the ResNet18 architecture.]{
		\includegraphics[width=0.7\textwidth]{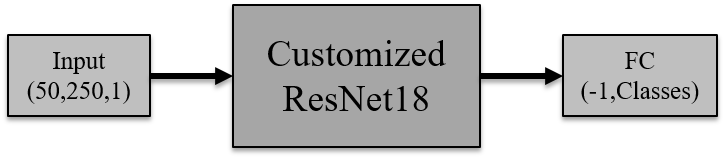}}
	\caption{Block diagram of the Model architectures}
	\label{fig:vggarc}
\end{figure*}

CNN has been proved to be efficient in signal processing. It can outperform classic methods in signal processing such as image and speech. Two different CNN architectures are proposed here. The first model is inspired by the VGG network
\cite{simonyan2014very}, and the second model is based on ResNet18 architecture
\cite{he2016deep}
which has a residual block in its structure.
These models are originally designed for image processing. In this study, they are employed for time series or one-dimensional signal processing.
Hence, some modifications are made to original architectures to fit desired conditions as below:
\begin{itemize}
	\item
	\textbf{Multi-channel inputs:}
	The original networks are designed for image processing which usually require three input channels as RGB.
	Here, the input channels are increased to 50, where all channels in the input contain traces related to a specific instruction from different programs and run times. This means that the pre-target and post-target instructions with all register values are different between channels and the only common part among these channels for one input is the target instruction. This forces network to extract features that distinguish the target instruction and does not get biased on other parts.
	\item
	\textbf{1D convolutions:}
	Both architectures employ 2D convolutions to extract spatial features from images. However, in our study 1D convolutions are used because our intention is to get more attention on some specific frequencies to distinguish different instructions. Furthermore, the tests illustrate that using the spatial information among multiple frequencies misleads the network to learn undesired features, leading to low accuracy.
	\item
	\textbf{SELU as activation function:}
	We alter the activation functions from rectified linear unit (ReLU) to scaled exponential linear units (SELU)
	\cite{klambauer2017self} since SELU does not completely discard the negative values. This results in higher accuracy during instruction recognition in the reverse engineering on our dataset.
\end{itemize}
By performing these modifications to original architectures, the first proposed model is shown in Figure
\ref{fig:vggarc}(a). The model is called M1 for the rest of this paper. M1 has a simple architecture, and its layers are limited. Based on ResNet18 architecture, another model called M2 (Figure
\ref{fig:vggarc}(b)) is proposed which is deeper than M1. This model includes the same architecture of the original ResNet18 with respect to the three mentioned modifications. These modifications make it suitable for our purpose, so that by using the residual blocks, the effect of a deep CNN is more observable on the dataset.

\subsection{Probability distribution estimation}
It is mentioned that the ARCSim generates entirely random instructions in recording the dataset. It should be noted that the instruction flow is not random in real programs at all, but it is based on some statistical distributions.
Estimating or even learning these distributions entirely or partially could assist reverse engineering models where the target device is a black box.
Hence, the proposed model in this section is employed to estimate these probability distributions and predict the next instruction.
Here, the number of previous instructions is significant because the first $q$ instructions are not predictable due to absence of initialize instructions.
LSTM networks are known for their ability to learn temporal features because of its memory. Suppose an attacker knows about the type of code running on the target device. In that case, he/she can use this information to estimate the probability distributions and eliminate the instructions with zero probability from training set for the CNN models.
The proposed instruction probability predictor based on LSTM is inspired by the hidden Markov model (HMM) presented in
\cite{eisenbarth2010building},
which aims to model the microprocessor as a state machine.
This analysis could determine the following instruction regarding the current instructions, specially could help CNN when the noise level is high and thus the confidence of its prediction is low. The structure of the LSTM model developed in our study is shown in Table.
\ref{tab:rnn}.
\begin{table}[ht]
	\begin{center}
		\caption{LSTM architecture}
		\label{tab:rnn}
		\begin{tabular}{|c|c|}
			\hline
			Layer & LSTM \\
			\hline
			Input & previous instruction labels \\
			\hline
			Layer 1 & 128 neurons \\
			\hline
			Layer 2 & 128 neurons \\
			\hline
			FC & (-1, class numbers) \\
			\hline
		\end{tabular}
	\end{center}
\end{table}

\section{Experimental results}\label{Exper}
This section includes the implementation procedure and the results for two proposed models.
We have trained and evaluated all proposed models on Nvidia GTX 1080 Ti.

\subsection{CNN for reverse engineering}
Two different models are explained in section
\ref{classification}
in order to trace classification.
In the training procedure, Adam optimizer and Cross-Entropy loss function are used. The learning rate is set to 0.0001, and after 40 epochs the learning rate is reduced to 0.00001.
The training is stopped at epoch 60 to avoid model overfitting.
Two attack scenarios are designed for reverse engineering using these two models as below:
\begin{itemize}
	\item
	\textbf{Sole scenario:}
	This method employs one model to classify all traces.
	\item
	\textbf{Hierarchical scenario:}
	A hierarchical attack contains two phases. First, instructions classified regarding their functionalities. Then, another model assigns a label for instructions of each group. This type of attack is inspired by
	\cite{park2018power}, however, some modifications are done here.
\end{itemize}
Both M1 and M2 are employed to deploy such attacks in both scenarios on datasets and the results are reported in Table
\ref{tab:CNNres}.
The first dataset has 45 classes from one-cycle instructions while the second dataset has 96 classes. Thus, the accuracy is higher in the first dataset compared to the second dataset. Besides, using the hierarchical scenario helps the network to divide the classification problem into two phases which results in higher accuracy.
However, reported accuracy values in Table
\ref{tab:CNNres}
are not comparable for both scenarios due to the existence of two phases in hierarchical scenario which have different probability weights.
Accordingly, an instruction-based table is available in the appendix A to determine the exact recognition accuracy using both scenarios for both datasets.
\begin{table*}[t]
	\begin{center}
		\caption{The classification accuracy of leakages for M1 and M2}
		\label{tab:CNNres}
		\begin{tabular}{|c|c|c|c|c|}
			\hline
			Model & \multicolumn{2}{c}{Scenario} & Dataset \#1 & Dataset \#2 \\
			\hline
			\multirow{4}{*}{M1} & \multicolumn{2}{c}{Sole} & 76.19\% & 49.37\% \\
			\cline{2-5}
			& \multirow{3}{*}{Hierarchical} & Grouping & 98.81\% & 95.13\% \\
			\cline{3-5}
			& & Worst/Best & 61.35\%/92.52\% & 44.64\%/98.68\% \\
			\hline
			\multirow{4}{*}{M2} & \multicolumn{2}{c}{Sole} & 78.27\% & 54.58\% \\
			\cline{2-5}
			& \multirow{3}{*}{Hierarchical} & Grouping & 99.02\% & 95.68\% \\
			\cline{3-5}
			& & Worst/Best & 42.10\%/92.16\% & 49.18\%/94.08\% \\
			\hline
		\end{tabular}
	\end{center}
\end{table*}

Among proposed methods, M2 outperforms M1 because of the residual blocks in the structure of M2.
After applying the proposed methods on datasets, two of the most known attacks proposed in
\cite{chari2002template}
and
\cite{park2018power}
are applied on datasets to compare with our proposed methods. Other existing methods usually include a small part of the instruction set which are not appropriate for comparison. PCA and LDA are combined with TA to evaluate their efficiencies in the reverse engineering using side channel information. One important implementation issue about TA is that our traces have 200 sample points, which causes computational problems in calculating the determinant of the sample matrix. Thus, one, 10, and 20 features are selected in three different experiments to solve this problem. These features are chosen at the points where traces between two classes have the maximum mean difference. The results are reported in Table \ref{tab:compare}.

\begin{table}[t]
	\begin{center}
		\caption{Comparison of the proposed method (M2 model) with other existing methods}
		\label{tab:compare}
		\begin{tabular}{|c|c|c|}
			\hline
			Method & Feature Points & Mean Accuracy \\
			\hline
			\multirow{4}{*}{TA} & 1 & 23.21\% \\
			\cline{2-3}
			& 10 & 21.08\% \\
			\cline{2-3}
			& 20 & 20.57\% \\
			\hline
			\multirow{4}{*}{PCA + TA} & 1 & 24.77\% \\
			\cline{2-3}
			& 10 & 22.15\% \\
			\cline{2-3}
			& 20 & 21.80\% \\
			\hline
			\multirow{4}{*}{LDA + TA} & 1 & 26.32\% \\
			\cline{2-3}
			& 10 & 25.54\% \\
			\cline{2-3}
			& 20 & 23.76\% \\
			\hline
			KL Div. + QDA \cite{park2018power} & 200 & 29.17\% \\
			\hline
			\textbf{Ours} & 200 & \textbf{54.58\%} \\
			\hline
		\end{tabular}
	\end{center}
\end{table}

\subsection{Instruction frequency analysis using LSTM}
The LSTM model is proposed for instruction frequency analysis and the probability estimation of instruction sequences. The dataset should not contain random instructions for this purpose. Since these models aim to extract the instruction flow of a real program, cryptographic algorithms are chosen as the target program to reverse. The provided training set includes eight cryptographic functions from AVR-crypto-lib
which has a standard implementation.
The customized version of simavr is employed to simulate the selected functions and export the label files. Then, some instructions are defined as the initial instructions for the network, which is the penalty for LSTM model.
In our experiments, initial instructions are set two, five, and 10 respectively to test the capability of model. Results are shown in Table
\ref{tab:RNNres}.
As expected, the network obtains more information from previous instructions by increasing the number of initial instructions. In fact, more instructions in the queue, higher prediction accuracy is achieved. However, the purpose of this network is to assist M1 or M2 model in a real attack scenario to increase the final accuracy. In this way, most of initial instructions will be determined by M1 or M2 and LSTM model will be used in the cases in which M1 or M2 could not reach a high confidence classification for a trace.
\begin{table}[ht]
	\begin{center}
		\caption{Frequency analysis results using LSTM (The accuracy of instruction prediction)}
		\label{tab:RNNres}
		\begin{tabular}{|c|c|c|c|}
			\hline
			\multirow{2}{*}{Model} & \multicolumn{3}{|c|}{Number of Initial Instructions} \\
			\cline{2-4}
			& 2 & 5 & 10 \\
			\hline
			LSTM & 67.34\% & 89.03\% & 94.39\% \\
			\hline
		\end{tabular}
	\end{center}
\end{table}

\section{Discussion}
Using CNN for side-channel trace processing helps the attacker to use unaligned data. As our four different types of instructions indicate, the target instructions are not aligned in different types. But, the proposed architecture can distinguish them in the attack phase. As a consequence, our proposed model shows that CNN is resistant to misalignment of signals.

M1 and M2 models are employed to analyze the leakage of each instruction. Referring to our results (appendix B), both the status register's bit instructions and arithmetic instructions are the two groups with fewer side-channel leakages. Thus, it is crucial to have more data to reduce the noise as much as possible. However, we suggest using the LSTM model alongside the M1 or M2 model to increase the confidence of output.
The LSTM model can be used in two different phases. First, in the profiling attacks, the attacker does have extra information about the target processor. Thus, he can use this prior knowledge to limit the training dataset of the M1 and M2 models. For example, suppose the attacker knows that the target processor is executing a cryptographic algorithm. In that case, he can extract the probability of the instructions and then eliminate the instructions with zero probability from the training dataset. Consequently, the M1 or M2 model will be trained only on the possible instructions.
Second, the LSTM model can be employed in instruction classification after M1 or M2. In this case, when the M1 or M2 model has more than one label assigned to a certain leakage, the LSTM model can be used to determine the correct label between the assigned labels by analyzing the previous instructions.

\section{Conclusion}\label{Con}
In this paper, we proposed a new mechanism for random dataset generation for disassembly attacks using side-channel information. The ARCSim was developed using python for instruction generation, simulation, and exporting labels file for profiling attacks. Later, our customized deep models are trained using the gathered dataset, and the results are compared to the three state-of-the-art side-channel attacks using machine learning. The practical results have shown the advantage of CNNs over other attacks in the feature extraction from low resolution data. Furthermore, a table including exact recognition accuracy per instruction in the hierarchical scenario is reported in the appendix B,
which could be referred to determine the leakage of each instruction. Referring to such table, we could specify the prediction confidence for each instruction in real attacks.
The dataset generated during the current study are available from the corresponding author on reasonable request.

By extending this work to more complex architectures, the amount of leakage could be determined, which could be helpful in designing new architectures to prevent these leakages and improve the security level. Another approach is to use attention-based neural networks to have better feature extraction. Furthermore, employing information-theoretic feature extractors in preprocessing could help the network training and will reduce the computational cost.

\ifCLASSOPTIONcaptionsoff
  \newpage
\fi



\bibliographystyle{IEEEtran}
%

\bibliography{bibliography.bib}




%








\end{document}